\documentclass[journal=jacsat,manuscript=article]{achemso}
\usepackage{graphicx}
\usepackage{float}
\usepackage{dcolumn}
\usepackage{bm}
\usepackage{setspace}
\usepackage{subcaption}
\usepackage{mathrsfs}
\usepackage{mathtools}
\usepackage{balance}
\usepackage{helvet}
\usepackage{xfrac}
\usepackage[dvipsnames,svgnames,x11names]{xcolor}

\usepackage[USenglish]{babel}
\usepackage{setspace}
\usepackage[most]{tcolorbox}
\setkeys{acs}{maxauthors = 0} 

\author{James M. Goff}
\affiliation{Department of Materials Science and Engineering, Materials Research Institute, The Pennsylvania State University, University Park, PA 16802, USA}
\email{jmg670@psu.edu}
\author{Francisco Marques dos Santos Vieira}
\affiliation{Department of Materials Science and Engineering, Materials Research Institute, The Pennsylvania State University, University Park, PA 16802, USA}
\author{Nathan D. Keilbart}
\affiliation{Lawrence Livermore National Laboratory, Livermore, California 94550, USA}
\alsoaffiliation{Department of Materials Science and Engineering, Materials Research Institute, The Pennsylvania State University, University Park, PA 16802, USA}
\author{Yasuaki Okada}
\affiliation{Murata Manufacturing Co., Ltd., 1-10-1 Higashikotari, Nagaokakyo-shi, Kyoto 617-8555, Japan}
\author{Ismaila Dabo,}
\affiliation{Department of Materials Science and Engineering, Materials Research Institute, The Pennsylvania State University, University Park, PA 16802, USA}
\alsoaffiliation{Penn State Institutes of Energy and the Environment, The Pennsylvania State University, University Park, PA 16802, USA}

\title{Predicting the pseudocapacitive windows for MXene electrodes with voltage-dependent cluster expansion models}

\begin{document}

\begin{abstract}
\normalsize
\singlespacing
MXene transition-metal carbides and nitrides are of growing interest for energy storage applications. These compounds are especially promising for use as pseudocapacitive electrodes due to their ability to convert energy electrochemically at fast rates. Using voltage-dependent cluster expansion models, we predict the charge storage performance of MXene pseudocapacitors for a range of electrode compositions. $M_3C_2O_2$ electrodes based on group-VI transition metals have up to 80 percent larger areal energy densities than prototypical titanium-based ( e.g.  $Ti_3C_2O_2$ ) MXene electrodes. We attribute this high pseudocapacitance to the Faradaic voltage windows of group-VI MXene electrodes, which are predicted to be 1.2 to 1.8 times larger than those of titanium-based MXenes. The size of the pseudocapacitive voltage window increases with the range of oxidation states that is accessible to the MXene transition metals. By similar mechanisms, the presence of multiple ions in the solvent ( $Li^+$ and $H^+$ ) leads to sharp changes in the transition-metal oxidation states and can significantly increase the charge capacity of MXene pseudocapacitors.
\end{abstract}

\begin{figure}
    \centering
    \includegraphics[width=\textwidth]{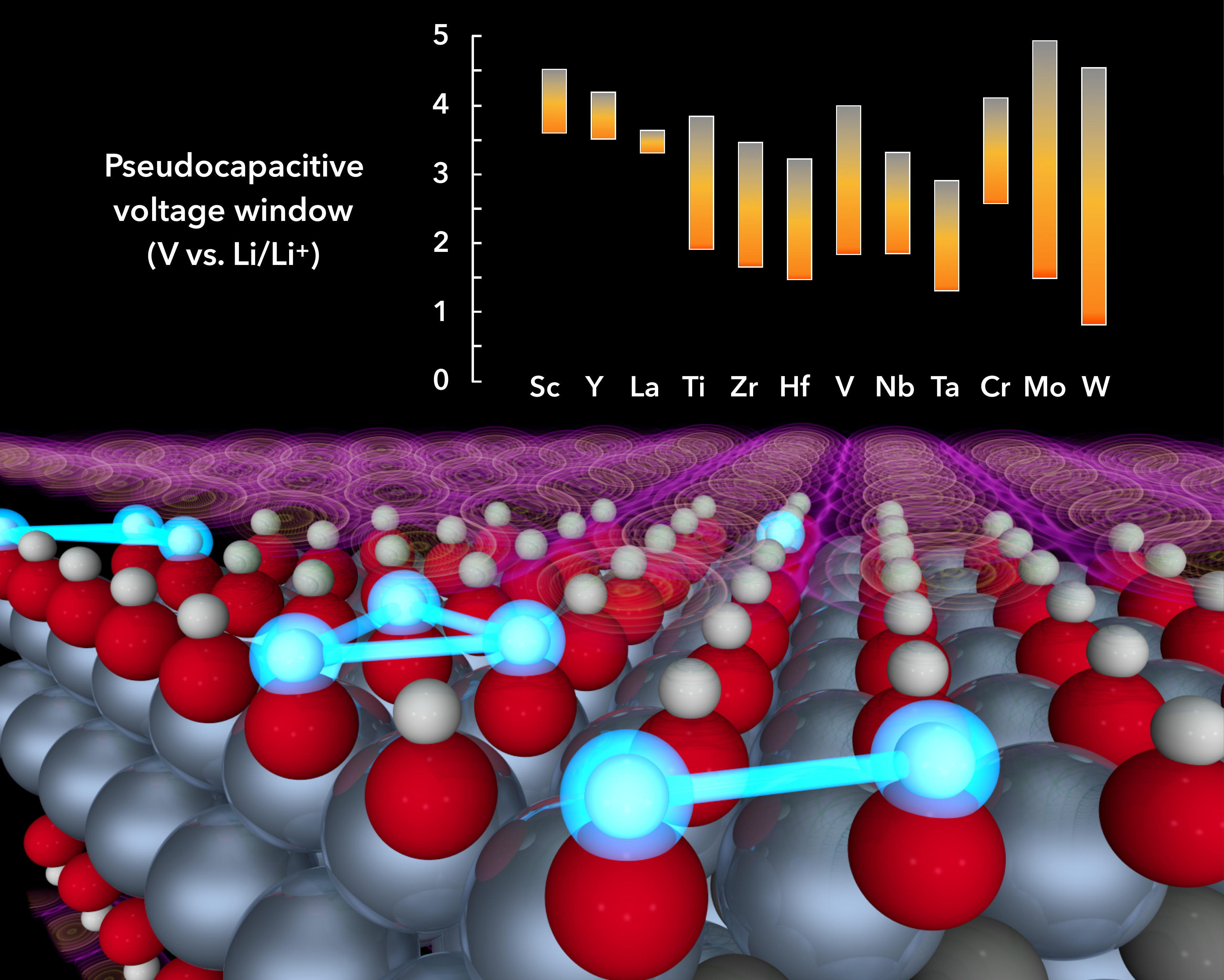}
    \caption*{Pseudocapacitive voltage windows predicted with voltage-dependent cluster expansions of MXene electrode surfaces }
    \label{fig:TOC}
\end{figure}

\clearpage

\section{Introduction}

\begin{figure}[h]
    \centering
    \includegraphics[width=\textwidth]{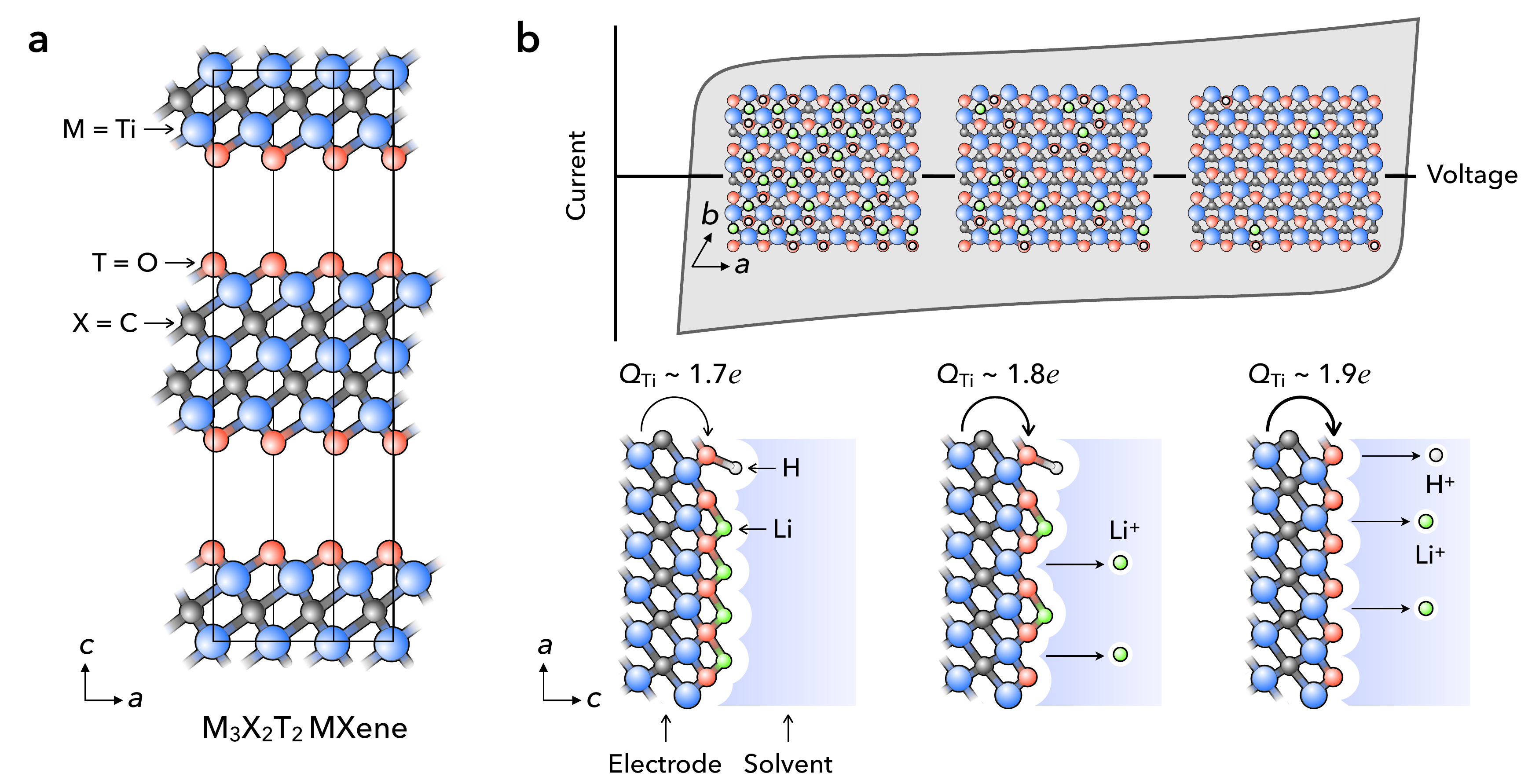}
    \caption{(a) Hexagonal crystal structure of the prototypical $\rm{Ti_3C_2O_2}$ MXene. (b) Idealized pseudocapacitive current-voltage response with corresponding ion adsorption states with representative structures (viewed along the $c$ axis) shown as insets (top). As the ion coverage decreases with the electrode voltage, the oxidation state of the transition metal increases (bottom). The nearly continuous set of redox reactions allowed by MXenes is responsible for the pseudocapacitive response. }
    \label{fig:MXene_general}
\end{figure}

MXenes have garnered attention for a range of applications, including catalysis, electronics, gas sensing, and energy storage.\cite{okubo_mxene_2018,seh_two-dimensional_2016,gogotsi_rise_2019} Their layered structures (Fig.~\ref{fig:MXene_general}a) and their electronic properties make them excellent candidates for pseudocapacitor and battery electrodes.\cite{lukatskaya_cation_2013,naguib_two-dimensional_2011,naguib_mxene_2012, er_ti3c2_2014} Pseudocapacitors are energy storage devices exhibiting higher energy densities than supercapacitors and higher power densities than batteries, which can be used in hybrid electrochemical systems to achieve electrical characteristics beyond the nominal performance of devices operating {\it via} a single charge storage mechanism ({\it e.g.} bulk intercalation or double-layer polarization).\cite{choudhary_asymmetric_2017} They are of practical relevance to high-power storage technologies for regenerative braking on trains, power electronics, and heavy machinery, among other potential uses.\cite{lukatskaya_probing_2015,wang_pseudocapacitance_2015}

Pseudocapacitors store charge by reversible redox reactions occurring near the surface of the material, and have characteristic capacitor-like current--voltage responses, as depicted schematically in Fig.~\ref{fig:MXene_general}b. They differ from electrochemical double-layer (EDL) capacitors in their charge storage mechanisms; the primary process of energy storage in pseudocapacitors is Faradaic charging---{\it i.e} charge accumulation due to near-surface chemical reactions. As evidenced by this and other studies, pseudocapacitors are characterized by a nearly continuous set of closely related redox states along the pseudocapacitive voltage window of the material.\cite{okubo_mxene_2018, lukatskaya_probing_2015,li_achieving_2017,fleischmann2020pseudocapacitance}  These different redox states can be induced in the transition metals by the adsorption of ions (H, Li, and Na, for example). Hence, MXene compositions that exhibit larger ranges of transition-metal oxidation states have wider Faradaic voltage windows. In this work we present predictions of the voltage windows for MXene pseudocapacitor electrodes as a function of composition. We highlight the effects of transition-metal valence and oxidation state on the energy capacity.

While ruthenia is a pseudocapacitive material with promising energy density, it has found limited commercial applications due to its high cost.\cite{zheng_hydrous_1995, noauthor_ruthenium_nodate} MXene materials, M$_{n+1}$X$_{n}$T$_2$, are exempt from this constraint---being composed of earth-abundant carbon (X = C, N) and oxygen (T = O, OH, F), and of relatively inexpensive transition-metal elements (M). These transition metals provide a wide compositional space, justifying the development of computational descriptors to rapidly assess pseudocapacitive performance.\cite{tan_high-throughput_2017,anasori_2d_2017} Electronic and thermodynamic descriptors have been used to predict trends in energy storage for MXene materials and have often been motivated by experimental observations. The oxidation state of the transition metal is closely related to the pseudocapacitive performance of MXenes.\cite{lukatskaya_probing_2015,zhan_understanding_2018} Related quantities such as the Bader charge of the transition metals, and the Fermi energy shift upon adsorption of hydrogen have been used to describe charge storage trends.\cite{zhan_computational_2019} Here, we aim to make connections between charge transfer descriptors (such as the transition-metal valence or Fermi energy shifts) and the active Faradaic voltage windows of MXene electrodes.

A number of models have been used to predict the energy storage performance of pseudocapacitive electrodes.\cite{ozolins_ruthenia-based_2013} Trends in Faradaic capacitance can be estimated for different MXene compositions using density-functional theory (DFT).\cite{zhan_understanding_2018} Local averaging methods can also be used to predict equilibrium charge--discharge curves.\cite{keilbart_quantum-continuum_2017} These DFT models can provide electronic descriptors, such as shifts in the potential of zero charge or metal oxidation states, which reflect the ability to transfer electronic charge between the adsorbed species and the pseudocapacitive surface.\cite{okada_mxene_2019} While the calculation of thermodynamic quantities over a range of adsorbate coverages can be generally accomplished with high accuracy in few-atom unit cells, the stability of larger surfaces with a high degree of configurational disorder at finite temperatures is rarely addressed. Additionally, DFT-based models generally provide an incomplete representations of lateral adsorbate interactions due to the limited length scales accessible to electronic-structure calculations. Adsorption phenomena at low and high voltages are key to predicting the active voltage windows, but are not accurately captured using DFT calculations alone.

A model that describes extended MXene electrodes at finite temperatures is needed to predict charge--discharge characteristics in the low- and high-voltage regimes. This can be accomplished with a two-dimensional surface cluster expansion model on the adsorbate sublattice(s) that contribute to the Faradaic charging. In the case of MXene pseudocapacitors, protic and aprotic lithium-containing solvents can be used to achieve excellent pseudocapacitive performance. By training a cluster expansion model with a dataset of DFT calculations that include an implicit solvent representation combined with double-layer charging, the effects of solvent and voltage can be accounted for along extended surfaces. The inclusion of these environmental conditions allows for a more realistic representation of the solid--liquid interface that approaches the accuracy of DFT calculations in some cases. This computational methodology is described and applied below.

The pseudocapacitive response of MXene electrodes is highly dependent on the solvent environment. For MXenes in aqueous electrolytes, intercalated water facilitates fast proton transfer to the electrode resulting in Faradaic charging for proton-driven pseudocapacitance.\cite{lukatskaya_ultra-high-rate_2017} For other ions such as Li$^+$ and Na$^+$, it was demonstrated that whether or not the ions were solvated leads to an electrochemical double-layer or pseudocapacitive response.\cite{gao2020tracking} In fact, different solvents facilitate the desolvation of participating ions.\cite{wang2019influences} Wang {\it et al.} have shown that pseudocapacitance at MXene electrodes in propylene carbonate solvents is dominated by charging processes involving desolvated ions. Using representative implicit solvent calculations, a clear distinction between electrochemical double-layer and Faradaic charging may be drawn.\cite{ando2020capacitive} We generate models in the regime that the response is dominated by near-surface charging processes; these also serve as first-order approximations of intercalation-based charging responses in the absence of inter-layer interactions. In reality, the overall electrochemical response will be the result of some combination of all charging processes. The predicted pseudocapacitve responses of MXene surfaces in this work can help determine which mechanisms contribute to charging, and may guide further experimentation and modeling efforts to understand and optimize MXene pseudocapacitors.

\section{Computational theory}

To model the adsorption mechanisms that contribute to charge storage, we begin by examining elementary surface processes. The charge stored per formula unit due to Faradaic charging can be calculated directly based on the coverage of a given species assuming integer charge transfer from the adsorbates:
\begin{equation}
    Q^{\rm{Faradaic}} = \sum_i N_i z_i
    \label{eq:faradaic}
\end{equation} 
where the sum runs over all adsorbing species, $i$. The product of the number of an adsorbing ion, $N_i$, and the respective valency of the adsorbing ion, $z_i$, gives the overall Faradaic charge stored. The equilibrium coverages and reversible charge-discharge curves can be determined {\it via} thermodynamic analysis.

In this approach, we first consider the adsorption of cations from the solution. Specifically, we aim to model electrodes in contact with lithium- and/or proton-containing solvents:
\begin{equation}
\begin{aligned}
   & * + {\rm H}^+ + e^- \longrightarrow {\rm H}^* \\
   & * + {\rm Li}^+ + e^- \longrightarrow {\rm Li}^*
\end{aligned}
\label{eq:adsorption_rxns}
\end{equation}
where $*$ indicates an adsorption site. The associated adsorption energies are given as
\begin{equation}
\begin{aligned}
&\Delta F ({\rm H}^*) = E({\rm H}^*) - E_{\rm MXene} - N_{\rm H} \frac{E_{\rm{H_2}}}{2}  \\
&    \Delta F ({\rm Li}^*) = E({\rm Li}^*) - E_{\rm MXene} - N_{{\rm Li}}E_{{\rm Li}} 
\end{aligned}
\label{eq:adsorption_energies_vac}
\end{equation}
and can be calculated from first principles. These adsorption energies vary with electrochemical conditions. We simulate these conditions using continuum embedding methods, which will be discussed below. In simulated electrochemical conditions, the free energies in Eq.~\eqref{eq:adsorption_energies_vac} should be modified by including (electro)chemical potentials:
\begin{equation}
\begin{aligned}
   &  \Delta F ({\rm H}^*)^{\rm solv} = E({\rm H}^*) - E_{\rm MXene} - N_{\rm H}\big( \mu_{{\rm H}^+} + \mu_{e^-} \big)  \\
    & \Delta F ({\rm Li}^*)^{\rm solv} = E({\rm Li}^*) - E_{\rm MXene} - N_{{\rm Li}}\big( \mu_{{\rm Li}^+} + \mu_{e^-} \big) 
\end{aligned}
\label{eq:adsorption_energies}
\end{equation}
where $\mu_{{\rm H}^+}$,  $\mu_{{\rm Li}^+}$, and $\mu_{e^-}$ are the electrochemical potentials of hydrogen ions, lithium ions, and the electrons in the electrode, respectively. 
These chemical potentials are evaluated as
\begin{equation}
\begin{aligned}
& \mu_{{\rm H}^+} = \frac{1}{2}E({\rm H}_2) +k_{\rm{B}}T {\rm ln}(a_{\rm H}) + e\phi_{{\rm H}|{\rm H}^+} \\
& \mu_{{\rm Li}^+} = E({\rm Li}) +k_{\rm{B}}T{\rm ln}(a_{\rm Li}) + e\phi_{\rm Li|Li^+} \\
& \mu_{e^-} = E_{\rm{F}}
\end{aligned}
\label{eq:chemical_potentials}
\end{equation}
using the free energies of hydrogen and lithium in their gaseous and metallic states, $E({\rm {H}}_2)$ and $E$(Li), respectively. The activity of the hydrogen and lithium in the solution, $a_{\rm {H}}$ and $a_{\rm {Li}}$, are taken to be 1 (under standard conditions). The standard reduction potentials of hydrogen and lithium, $\phi_{{\rm {H}}|{\rm {H}}^+}$ and $\phi_{{\rm {Li}}|{\rm {Li}}^+}$ on the absolute scale, multiplied by the fundamental charge $e$, is the chemical work to oxidize the adsorbates. The reduction potentials of hydrogen and lithium in water under standard conditions are used; these are 3.04 and 0.0 V on the lithium scale, respectively. The absolute electrochemical potential of the electron is equal to the Fermi energy, $E_{\rm{F}}$ (by definition). 

Depending on the electrode voltage, the equilibrium surface coverage---and the stored Faradaic charge---on the MXene surface will change. We account for this effect by expanding the adsorption energies in Eq.~\eqref{eq:adsorption_energies} with respect to the charge 
\begin{equation}
    \Delta F^{\rm{ads}} (q) =  \Delta F^{\rm{ads}} (0) + q\phi_{\rm pzc} + \frac{q^2}{2{\mathscr A} C_{\rm{dl}}} + ...  
    \label{eq:charge_expansion}
\end{equation}
where the charge on the surface is $q$, $\mathscr A$ is the surface area, and $C_{\rm dl}$ is the differential double layer capacitance of the MXene electrode. The terms of this expression can be calculated using continuum embedding models. The potential at which there is zero excess charge on the surface is the potential of zero charge, $\phi_{\rm pzc}$, which is given by $-E_{\rm{F}}/e$. A voltage-dependent adsorption energy can be obtained by truncating this expansion at second order and taking the Legendre transform with respect to charge:
\begin{equation}
    \Delta \mathscr{F}^{\rm{ads}} (\phi) =  \Delta F^{\rm{ads}} (q) -q\phi 
    \label{eq:voltage_expansion}
\end{equation}
where $\mathscr{F}^{\rm{ads}}$ is the transformed free energy and $\phi$ is the electrode voltage. In the calculation of this voltage-dependent adsorption energy, we note that the electron chemical potential in Eq.~\eqref{eq:adsorption_energies} is set by the electrode voltage $\mu_{e^-}=-e\phi$.

The prediction of voltage-dependent adsorption energies in Eq.~\eqref{eq:voltage_expansion} using DFT models is limited to the length scales accessible. This often results in coarse samplings of ion coverages, and models produced from these do not often capture longer range lateral interactions between adsorbates. Additionally, the convex hull is densely populated with adsorption configurations near the lowest energy ground state, as seen in Fig.~S2 of the Supporting Information. In order to accurately predict adsorption phenomena as a function of voltage, statistical ensemble models of extended surfaces are required. Cluster expansions and Monte Carlo simulations can be used to build such models.

\begin{figure}[H]
    \centering
    \includegraphics[width=\textwidth]{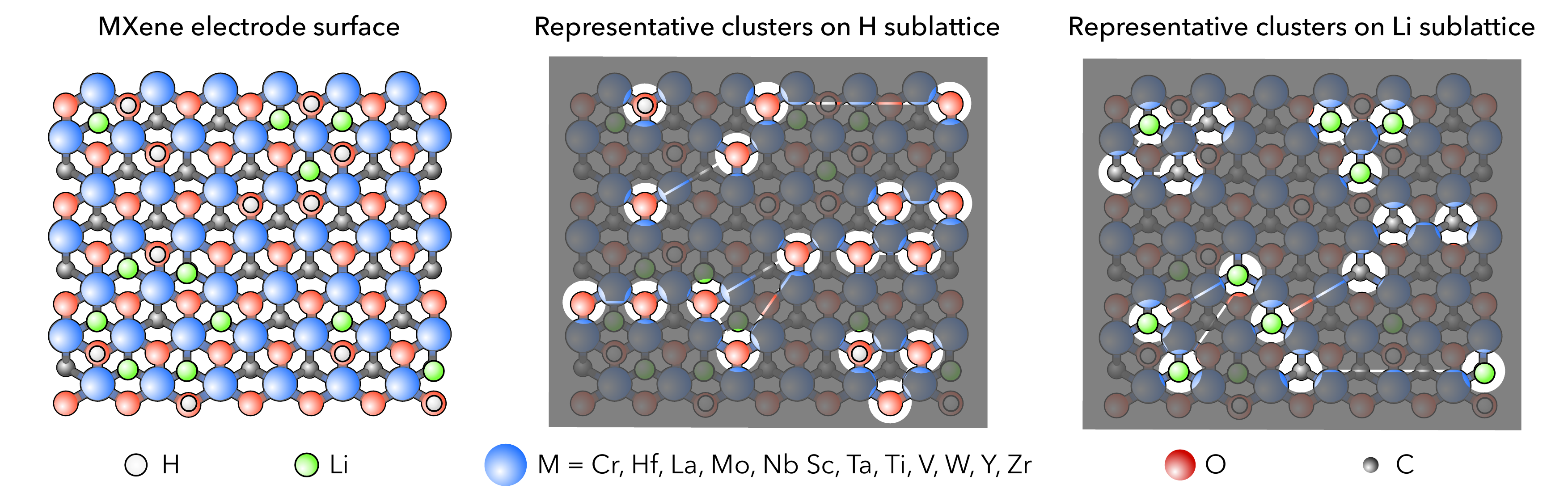}
    \caption{Construction of the cluster basis on the hexagonal MXene surface (left) for the hydrogen sublattice (center) and lithium sublattice (right).}
    \label{fig:clusters}
\end{figure}

The energies in Eq.~\eqref{eq:adsorption_energies} vary as a function of the coverage of the respective ions, and the nearly continuous set of adsorption states on the electrode surface contributes to the pseudocapacitive response described in Eq.~\eqref{eq:faradaic}. Therefore statistical ensemble representations of the electrodes and inclusion of lateral adsorbate interactions are important to accurately predict the pseudocapacitance. It is possible to represent scalar extensive functions of configuration as a linear expansion with the basis functions, following Sanchez and coworkers.\cite{sanchez_generalized_1984} The adsorption energy on a single  sublattice can be represented as an expansion in this spectral basis,
\begin{equation}
    F(\boldsymbol{\sigma}) = \sum_{\alpha} j_\alpha \Phi_\alpha(\boldsymbol{\sigma})
    \label{eq:ce_neutral}
\end{equation}
where $\boldsymbol{\sigma}$ is a vector of occupational variables giving the occupancy of each site on the sublattice ({\it e.g.}~--1 for vacancy and +1 for lithium ions). The sum in the expansion runs over clusters, $\alpha$, which are illustrated in Fig.~\ref{fig:clusters}. There is an expansion coefficient associated with each cluster, $j_\alpha$, and the spectral basis function for this cluster, $\Phi_\alpha(\boldsymbol{\sigma})$. This expansion can be used to predict the adsorption energy of configurations for extended surfaces that are more representative of electrode surfaces. In practice, the cluster basis functions in Eq.~\ref{eq:ce_neutral} are averaged over the crystal and the same expansion coefficient is used for symmetrically equivalent clusters.\cite{puchala_thermodynamics_2013}

We can also expand the voltage-dependent adsorption energy to obtain an energy expression for arbitrary adsorption configurations on a single adsorption sublattice as a function of voltage:\cite{weitzner_voltage-dependent_2017}
\begin{equation}
    \mathscr{F}(\boldsymbol{\sigma}, \phi) = \sum_{\alpha} J_\alpha(\phi)\, \Phi_\alpha(\boldsymbol{\sigma})
    \label{eq:ce_voltage}
\end{equation}
In this voltage-dependent cluster expansion, the coefficients are defined at each voltage. These can be obtained by training against a dataset of calculated voltage-dependent adsorption energies, as described in Eq.~\eqref{eq:voltage_expansion}. 

The co-adsorption of multiple ions cannot be modeled with Eq.~\eqref{eq:ce_neutral} and Eq.~\eqref{eq:ce_voltage} if the adsorption sites are on different, non-exchanging sublattices. Adsorption configurations on one sublattice need to be designated with a vector of spin variables, $\boldsymbol{\sigma}$. The adsorption configurations on the sublattice corresponding to the other sites need to be specified with another vector of spin variables, $\boldsymbol{\delta}$, and the multicomponent cluster expansion formalism developed by Tepesch, Garbulsky, and Ceder should be used.\cite{tepesch_model_1995} We adapt this approach to also include voltage dependence:
\begin{equation}
    \mathscr{F}(\boldsymbol{\sigma},\boldsymbol{\delta}, \phi) = \sum_{\alpha\beta} V_{\alpha\beta}(\phi)\,  \Theta_{\alpha\beta}(\boldsymbol{\sigma},\boldsymbol{\delta})
    \label{eq:ce_coads}
\end{equation}
This expansion is facilitated by the construction of basis functions of the coupled sublattices. These are constructed by taking the tensor product of the basis functions \{$\Phi_\alpha$\} for each sublattice forming the basis functions for the coupled sublattices, $\Theta_{\alpha\beta}(\boldsymbol{\sigma},\boldsymbol{\delta})$ ({\it  e.g.}~all possible products of clusters in Fig.~\ref{fig:clusters}). Thus there are two cluster indices, $\alpha$ and $\beta$. The expansion coefficients, $V_{\alpha\beta}(\phi)$, include effective cluster interactions (ECIs) from single-sublattice clusters as well as multiple-sublattice clusters. Using this formalism, a method for modeling the simultaneous adsorption of hydrogen and lithium is achieved. 

The expansion coefficients include the effects of solvent polarization and the diffuse double layer when fit to DFT calculations of surface cells in implicit solvents. The MXene electrode systems are embedded in a dielectric cavity and a smooth Poisson--Boltzmann density of countercharges set at a fixed distance of 3.5 \AA\ away from the MXene surface using a planar interface.\cite{andreussi_solvent-aware_2019} This interface function is chosen to reduce excessive screening between adsorbates. At the boundary, a planar cavity function for the continuum dielectric and implicit ion charges, $s(\mathbf{r})$, is used with the dielectric function, $\epsilon(\mathbf{r}) = \exp({\rm ln}(\epsilon_0)[1-s(\mathbf{r})])$, where  $\epsilon_0$ is the dielectric constant of the bulk solvent.\cite{nattino_continuum_2018} The dielectric was specified here as 12 for an approximate representation of solvent polarization in simple organic solvents. The charge density for the $i^{\rm th}$ ion, is given by the concentration in the solvent, $c_i$, multiplied by a Boltzmann weight. A sum over implicit ion types gives the total density, $\rho_{\rm{PB}}(\mathbf{r}) = \sum_i (1-s(\mathbf{r}))c_i \exp\left({-({z_i e \phi(\mathbf{r})}) / ({k_{\rm B} T})}\right)$, and contributions to the DFT energy are solved for self-consistently.\cite{nattino_continuum_2018} Examples of the resulting implicit Poisson--Boltzmann distribution of ions and bound polarization charges for the solvent are provided in Fig.~S1 of the Supporting Information.

\section{Computational methods}
Cluster expansion models were parameterized for $\rm{M_2CO_2}$ and $\rm{M_3C_2O_2}$ MXene electrodes where M is an early transition metal: Cr, Hf, La, Mo, Nb, Sc, Ta, Ti, V, W, Y, Zr. The adsorption of lithium and hydrogen on the MXene surfaces were considered independently to simulate lithium-containing and protic solvents. Single-ion adsorption models were made using Eq.~\eqref{eq:ce_voltage}. In protic, lithium-containing solvents, it is possible for lithium and hydrogen to co-adsorb. Hydrogen prefers the oxygen-top sites and lithium generally prefers the hexagonal closed-packed (hollow) site.

The cluster expansion models were fit using the Clusters Approach for Statistical Mechanics (CASM) code \cite{puchala_thermodynamics_2013}, with automated Python wrappers to incorporate the voltage dependence into the cluster expansions. All of the symmetrically distinct pair clusters within 3 adsorption sites, triplets within 3 adsorption sites, and quadruplet clusters within 2 adsorption sites were used in the cluster expansion models along with the identity and single site clusters. Discrete Chebychev site basis functions were used to construct the cluster basis functions. Optimal sets of expansion coefficients are determined at each voltage using $l_1$-regularized least squares fit against DFT data. The test errors of the fits are estimated with $k$-fold ($k$ = 10) cross validation along with cluster elimination as implemented in CASM. The test errors are reported in the Supporting Information in Fig.~S6.

The DFT training sets were comprised of 184 and 53 symmetrically distinct adsorption configurations for multiple-ion and single-ion adsorption systems, respectively. Semi-local DFT calculations were performed at the GGA level in Quantum ESPRESSO to calculate the energies of the structures in the fitting set.\cite{giannozzi_quantum_2009} Plane wave basis sets were used with a kinetic energy cutoff of 120 Ry, and norm-conserving pseudopotentials were employed to represent atomic cores.\cite{van_setten_pseudodojo_2018} The Brillouin zone was sampled in all cells using a (11/$n$ $\times$ 11/$m$ $\times$ 1) Monkhorst-pack {\bf k}-point mesh where $n$ and $m$ are the multiples of the primitive surface cell vectors. A cold smearing value of 0.01 Ry was used to smooth electronic occupations.\cite{marzari_thermal_1999} All fitting cells were embedded in a 15 \AA\ continuum dielectric and implicit ion distribution to represent the solvent (cf. the previous section).

The implicit ion distribution in our continuum solvent model allows for a DFT-level approximation of the differential diffuse layer capacitance. Explicit charge applied to the MXene electrode is screened by the ion distribution in the solvent region. The energy as a function of charge can be used to fit the quadratic form of Eq.~\eqref{eq:charge_expansion}, where the second-order term is related to the diffuse double-layer capacitance, $C_{\rm dl}$. This double layer capacitance varies with adsorption configuration with an average value of 8.6 $\mu$F/cm$^2$ in a test MXene composition, Mo$_3$C$_2$O$_2$. It was found that using a constant diffuse double layer capacitance, 10 $\mu$F/cm$^2$, close to the average yields similar results. A constant capacitance of 10 $\mu$F/cm$^2$ was used for all MXene compositions to evaluate voltage-dependent adsorption energies in Eq.~\eqref{eq:ce_voltage}. 

\begin{figure}[H]
    \centering
    \includegraphics[width=0.7\textwidth]{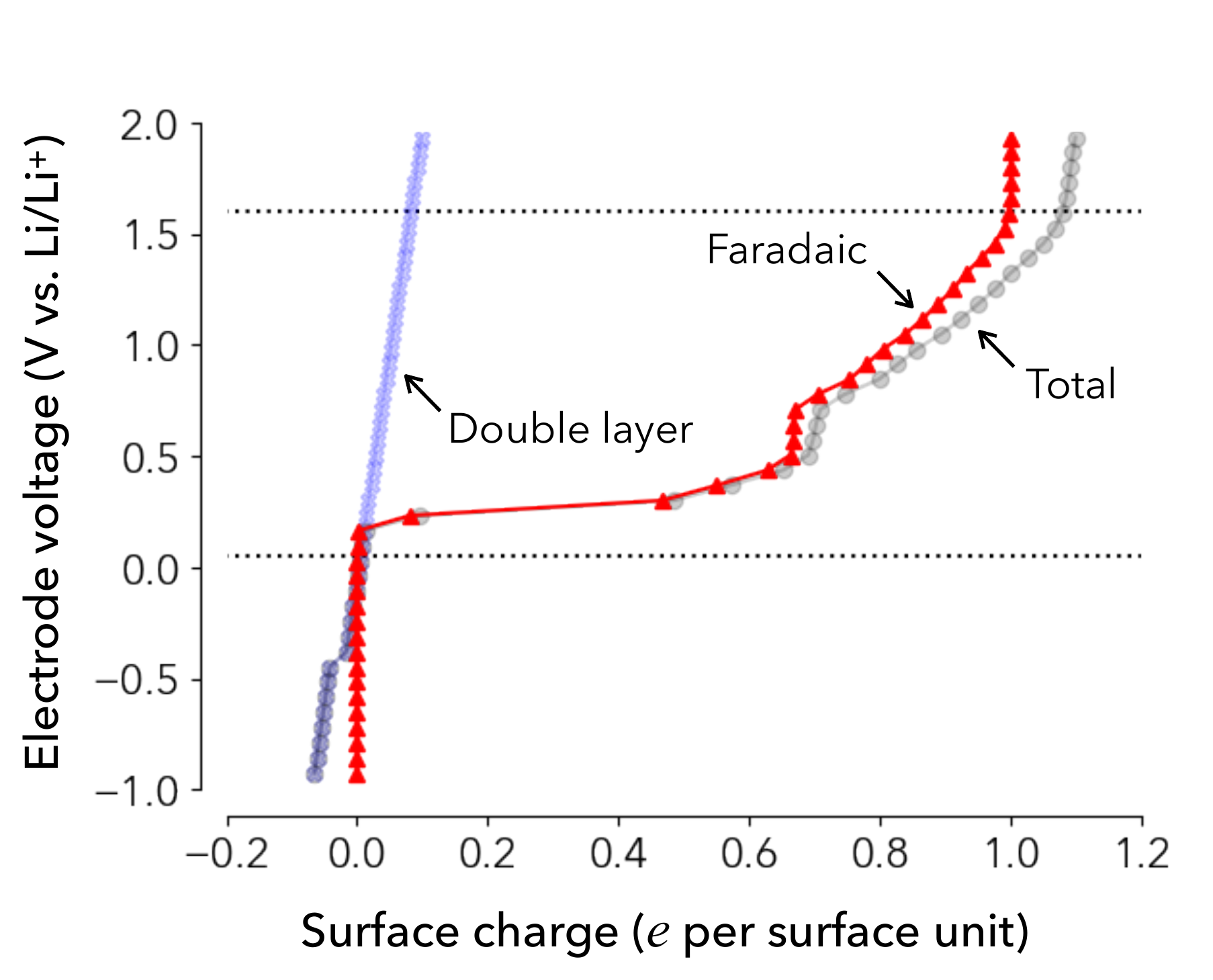}
    \caption{Predicted charge--voltage response of the a $\rm{Ti}_2\rm{C}\rm{O}_2$ MXene electrode from Monte Carlo simulations. The blue curve is the charge stored due to diffuse electrochemical double-layer charging, the red curve is the Faradaic charge stored, and the black curve is the net charge stored. The decoupled pseudocapactive charge storage response is used to predict the active voltage window of the electrode indicated by the horizontal lines. }
    \label{fig:isotherm}
\end{figure}

Monte Carlo simulations were performed using the CASM software package. Simulations for hydrogen or lithium ion adsorption on MXene electrodes were performed in the grand-canonical ensemble on 30 $\times$ 30 surface cells. In these simulations, the steady-state coverage of lithium or hydrogen adsorbates was determined at each voltage in 10 separate simulations and averaged; the resulting adsorption isotherms are reported in Figs. S7-S10 of the Supporting Information with an example highlighted for $\rm{Ti}_2\rm{C}\rm{O}_2$ in Fig.~\ref{fig:isotherm}. To obtain these isotherms, each Monte Carlo simulation was ran for 1,000 passes after the system had been equilibrated, and were initialized with a random adsorption configuration. The steady-state coverage of lithium ions as a function of voltage allows us to determine the active pseudocapacitive voltage window. This is the voltage window where the slope of the adsorption isotherm is non-zero and the ion coverage is between 0 and 1 monolayers. Additionally, the slopes of the isotherms are directly related to the pseudocapacitance of the MXene electrodes.

In lithium-containing protic solvents, the effect of hydrogen adsorption at small coverages can help extend the active voltage window for lithium. To predict the extent of lithium stabilization in the presence of other ions, we simulate lithium and hydrogen adsorption on MXene electrodes in low coverage regimes. These simulations were also performed on 30 $\times$ 30 surface cells to represent the electrode with a single Monte Carlo simulation (10,000 passes) at each voltage. The cluster expansions fit at each voltage include small cells with both hydrogen and lithium on the surface. Using the multilattice cluster expansion models in Eq.~\eqref{eq:ce_coads} for fixed composition Monte Carlo simulations, we predict the stabilization of lithium adsorption in the upper voltage limit. Solving for the voltages with zero formation energy in Eq.~\eqref{eq:voltage_expansion} in regimes with small lithium and small hydrogen ion adsorption can reveal the extent to which the introduction of hydrogen may favor lithium adsorption.

\section{Results and discussion}

\begin{figure}[H]
\centering
\includegraphics[width=0.86\textwidth]{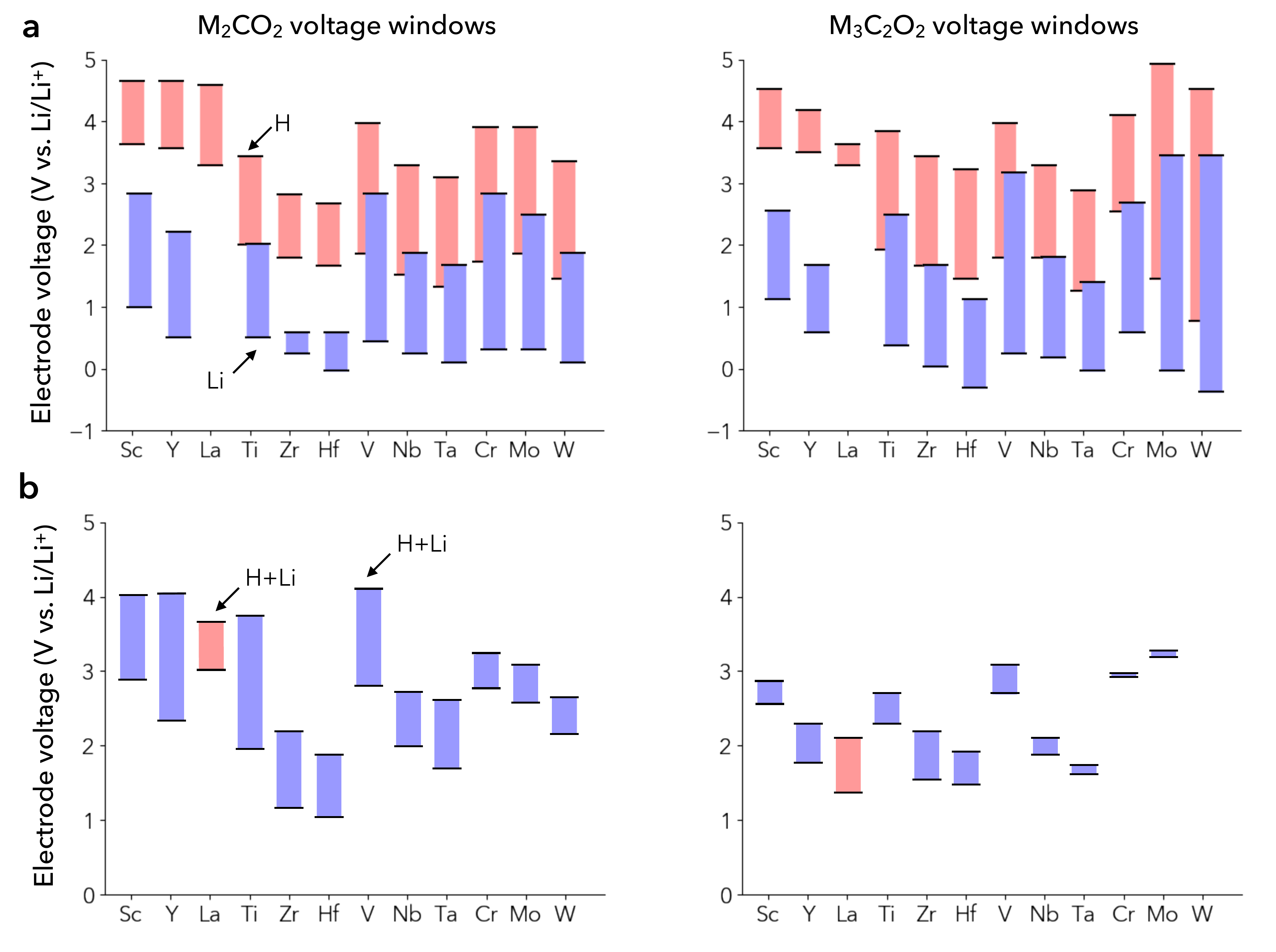}
\caption{ (a) Pseudocapacitive voltage windows for oxygen terminated MXenes determined from adsorption isotherms at fixed voltage. Lithium voltage windows are given in blue and hydrogen voltage windows in red.  (b) Average voltage window extension calculated by evaluating the average $\phi (\mathscr{F}=0)$ for a range of low H and Li coverages, giving the upper voltage limit. The lower voltage limit is $\phi (\mathscr{F}=0)$ predicted by Monte Carlo simulations without H present. }
\label{fig:volt_windows}
\end{figure}

The equilibrium charge--discharge curves (Fig.~\ref{fig:isotherm}) were obtained for all M$_3$C$_2$O$_2$ and M$_2$CO$_2$ early-transition-metal compositions. From these, the voltage windows with pseudocapacitive activity are calculated as the range of voltages where the derivative of the Faradaic charge with respect to voltage is non-zero and the coverage of the adsorbing ion is between but not equal to 0\% and 100\%. These voltage windows are compiled in Fig.~\ref{fig:volt_windows}a for the different MXene compositions. The Faradaic capacitances, were evaluated by taking the derivative of the Faradaic charge with respect to voltage over the active window and averaged. We note that an instantaneous capacitance could be obtained by a simple numerical derivative of this charge--voltage curve, but to be accurate, this would require a much finer numerical grid than is sampled. The average capacitances are reported in Table S1 of the Supporting Information along with the adsorption isotherms used to calculate them. The predicted isotherms are consistent with other first-principles based models of MXene electrode surfaces.\cite{zhan_understanding_2018, okada_mxene_2019} Relatively linear decreases in ion coverages as a function of voltage correspond to Faradaic charging and the characteristic rectangular cyclic voltammetry response for capacitors. The equilibrium charge/discharge curves can be contrasted with similar simulations of lithium-ion battery materials, which are associated with rapid decrease in Li content over a short voltage range, to show that many MXene compositions yield pseudocapacitive charge/discharge behavior.\cite{wolverton1998first} Pseudocapacitive responses and voltage windows were highlighted for the group-VI MXene electrode surfaces in simulated organic solvent environments. Improved accuracy is expected over existing models with the voltage-dependent cluster expansion method at high and low ion coverages because of the third-nearest neighbor lateral adsorbate interactions trained in the models.

The size of the voltage windows increases from left to right in the periodic table. The voltage windows tend to decrease with increasing period, except with the group-VI M$_3$C$2$ MXenes. An ideal pseudocapacitor has a constant Faradaic capacitance over a large voltage window while desorption over a small voltage range is more characteristic of a battery electrode. Thus, Zr and Hf are not suitable lithium-based pseudocapacitor candidates, but may perform better as lithium-ion battery electrodes. The active voltage windows for hydrogen are higher and closer to the standard redox potential of hydrogen. For some transition-metal compositions, the active voltage windows of hydrogen and lithium overlap, indicating that in protic, lithium-containing solvents there may be voltage ranges where both ions contribute to energy storage. We draw attention to the group-VI MXene compositions in Fig.~\ref{fig:volt_windows}a; these compositions have large active windows for both hydrogen and lithium pseudocapacitance. 

The pseudocapacitive windows of lithium and hydrogen overlap, suggesting that, at some potentials, there is simultaneous adsorption of hydrogen and lithium. While adsorption of large amounts of lithium and hydrogen were predicted to lead to significant reconstructions for all MXenes, small amounts of lithium and hydrogen, corresponding to the upper voltage lithium voltage window, are stable. The hydrogen partially stabilizes the adsorption of lithium at higher voltages. This observation is supported by fixed composition Monte Carlo simulations with small amounts of hydrogen $\theta_{\rm H} =$  \sfrac{1}{9}, \sfrac{2}{9}, and \sfrac{3}{9} and small amounts of lithium, $\theta_{\rm Li} =$ \sfrac{1}{10}. With these simulations, the voltage at which adsorption is stable can be obtained with Eq.~\eqref{eq:voltage_expansion} and the average extension of the Li voltage window can be evaluated for both M$_3$C$_2$ and M$_2$C MXenes. The average voltage window extension upon introduction of hydrogen is given for each MXene composition in Fig.~\ref{fig:volt_windows}b. The effect is larger for  M$_2$C MXenes and the voltage window tends to be larger for earlier transition metals. For the W$_3$C$_2$O$_2$ MXene, the voltage window extension due to multiple ion adsorption is negligible. Lithium adsorption by itself leads to reconstructions on lanthanum-based MXenes. Lithium adsorption is stabilized to a small extent with the introduction of hydrogen, and is indicated by the red bars in Fig.~\ref{fig:volt_windows}b. The stabilization of Li ion adsorption at high voltages is due to increased interactions between Li and the transition metals in the presence of H adsorbates. This observation can be explained by examining the density of states reported the Supporting Information, which reveal increased charge transfer to the MXene transition metal in the presence of multiple ion species.
    
\begin{figure}[H]
\centering
\includegraphics[width=\textwidth]{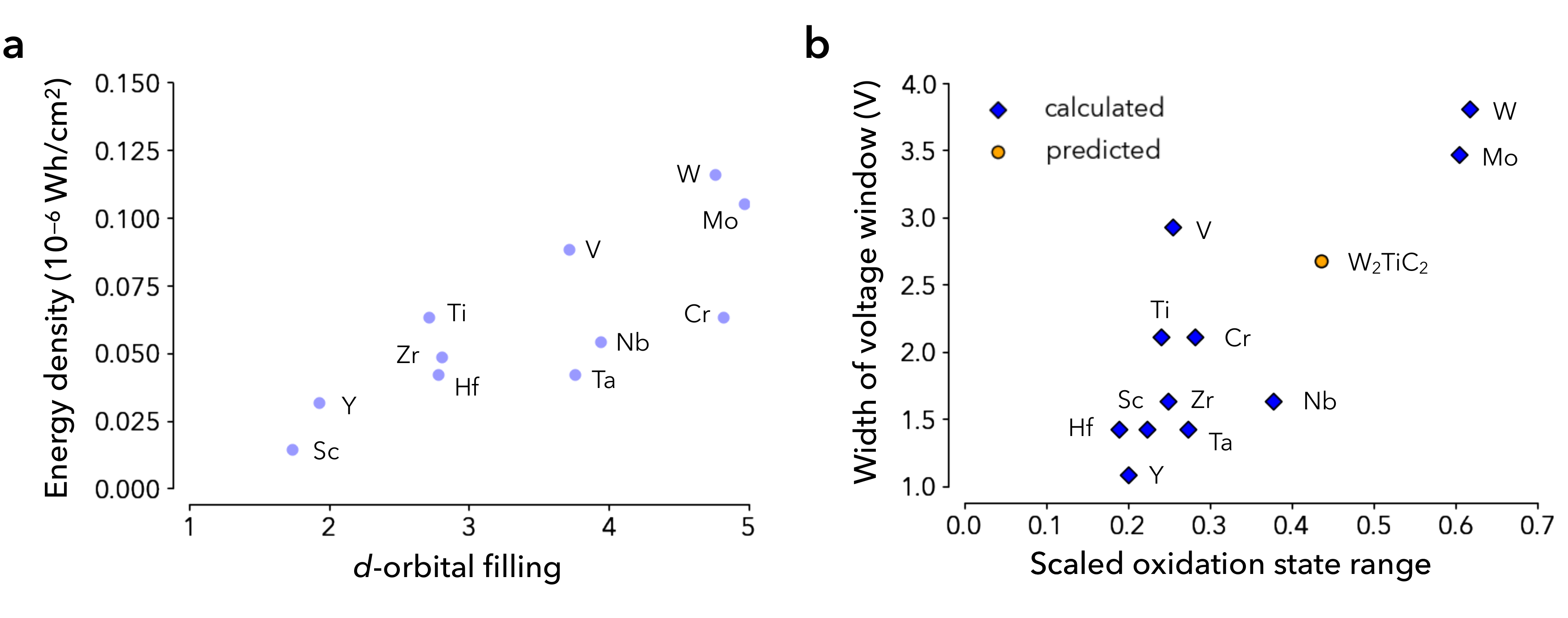}
\caption{(a) Pseudocapacitive energy density for lithium adsorption on M$_3$C$_2$O$_2$ MXene as a function of Lowdin \textit{d}-orbital population. The \textit{d}-orbital filling is calculated on MXene slabs and is proportional to the transition-metal valence. The energy density increases with \textit{d}-orbital filling and follows row and column periodic table trends. (b) The pseudocapacitive voltage window size increases with the scaled change in oxidation states during adsorption induced redox on MXene surfaces with the M$_3$C$_2$O$_2$ compositions.} 
\label{fig:M3C2_energies}
\end{figure}

The pseudocapacitive energy storage is calculated using the average Faradaic capacitances $C_{{\rm {Faradaic}}}$ obtained in Table S1 of the Supporting Information and the active voltage windows $\Delta V$ in Fig.~\ref{fig:isotherm} with the equation $E_{\rm{pseudo}} = \frac{1}{2}C_{{\rm Faradaic}}\Delta V^2$. We report the lithium pseudocapacitive energy densities for $\rm{M_3C_2O_2}$ MXenes in Fig.~\ref{fig:M3C2_energies}a in the presence of lithium-containing solvents. The trends for $\rm{M_2CO_2}$ MXenes and hydrogen-containing solvents are similar and are included in the Supporting Information. We note that these pseudocapacitive energies are decoupled from the electrochemical double-layer energy storage. These predictions of MXene energy storage capabilities show that the pseudocapacitive energy stored correlates with the valence of the transition metal. This trend can be explained by the fact that redox reactions involving the transition metal are primarily responsible for the pseudocapacitive response in MXenes. There are generally more oxidation states available for late-transition-metal (group-V and group-VI) MXenes, and this facilitates Faradaic charging from adsorbed ions and larger pseudocapacitive windows. We confirm this with relative ranges of oxidation states for M$_3$C$_2$O$_2$ MXenes in Fig.~\ref{fig:M3C2_energies}b. Larger ranges of oxidation states generally indicate larger pseudocapactive windows, but magnetic MXenes, V and Nb-based, follow more irregular trends. The V- and Nb-based MXenes deviate most from linear increases in voltage window size with oxidation state range; a thorough study of the effects of magnetism on pseudocapacitive charge storage would be beneficial for these systems.

It has been shown that the Bader charge is linearly proportional to experimental oxidation states in MXenes, but the constants of proportionality may be different from composition to composition.\cite{zhan_understanding_2018} For this reason, the oxidation state ranges in Fig.~\ref{fig:M3C2_energies}b were predicted as relative changes in Bader charge upon redox with Li$^+$. The changes in Bader charge for the surface transition metals, reported in the Supporting Information, were multiplied by a scaling factor, $f=\Delta \phi_{\rm{pzc}}({\rm M})/\Delta \phi_{\rm{pzc}}({\rm Ti})$. Other scaling factors could be used to describe relative ranges of oxidation states, but the ratio $f$ bears valuable physical meaning. It describes the change in the affinity of the MXene surface for charged species relative to another and is an important, reliable descriptor for the Faradaic capacity.\cite{zhan_computational_2019} The combination of these descriptors help describe the trend found with group-VI MXenes that when the MXene transition metals access larger ranges of oxidation states, larger windows of Faradaic charging are observed. We note that the group-VI $\rm{M_3C_2O_2}$ MXenes are not typically stable due to the transition-metal layer stacking preference. Group-VI metals prefer the ABAB stacking but possess ABCABC stacking in the $\rm{M_3C_2O_2}$ compositions.\cite{anasori_2d_2017} Late transition metals can be stabilized in $\rm{M_3C_2O_2}$ MXenes by forming ordered alloys with other transition metals such as Ti. We show in Fig.~\ref{fig:M3C2_energies}b that one such MXene composition, $\rm{W_2TiC_2O_2}$, accesses a wide range of oxidation states as well, and is expected to have a large pseudocapacitive voltage window.

The search for new MXene pseudocapacitor electrodes should consider MXene compositions that allow for large ranges of transition-metal oxidation states, and solvent environments that facilitate them. For different solvent, electrolyte, and electrode combinations, different charging processes will dominate the electrochemical response.\cite{gao2020tracking} The voltage-dependent cluster expansion models used in this work correctly represent MXene-electrode systems where surface charging dominates, highly exfoliated MXenes and systems where intercalation charging is expected to contribute less. Though our results serve as an approximate model of intercalation charging in the absence of inter-layer interactions, they also highlight systems where Li$^+$ intercalation charging effects are important and would benefit from further direct investigation, such as with heavy group-V $\rm{M_2C}$ MXenes given the small predicted voltage windows. 

Considering alloy MXenes as pseudocapacitive electrodes beyond the ${\rm{W_2TiC_2O_2}}$ composition will be important. The presence of other transition metals or X elements could induce varied oxidation states, depending on chemical ordering or lack thereof, that enhance pseudocapacitive energy storage through a combination of larger voltage windows or capacity.\cite{tan_high-throughput_2017} Increased pseudocapacitive voltage windows have been observed experimentally. It was confirmed by Vahidmohammadi {\it et al.} that $\rm{V_2CO_2}$ MXenes have larger pseudocapacitive voltage windows compared to the early-transition-metal or Ti MXene electrodes.\cite{vahidmohammadi2019assembling,okubo_mxene_2018} There are promising results for group VI (Mo-based) MXenes in similar conditions.\cite{tao2017two} The key effect of transition-metal valence and the range of induced oxidation states in the MXene transition metals on the MXene pseudocapacitance could be confirmed with x-ray adsorption spectroscopy experiments, but are qualitatively anticipated given the wider range of oxidation states allowed for vanadium relative to titanium.\cite{zhan_understanding_2018}

\section{Conclusion}

Voltage-dependent cluster expansion models were used to predict Faradaic charge storage in MXenes across various transition-metal compositions. We found that group-VI MXenes may be promising pseudocapacitor candidates due to large pseudocapacitive voltage windows that lead to high energy storage densities. The predicted pseudocapacitive voltage windows only describe voltages where the MXene electrodes will be active with Faradaic charging, but do not provide information about the breakdown voltage of the MXene electrode. These breakdown voltages will depend on the solvent, the ionic species, and the composition of the electrodes. A study of the electrochemical durability of MXene electrodes would show how much of the Faradaically active window is accessible for different solvent--electrode combinations. The Faradaic voltage windows for lithium adsorption on these MXenes may be accessible with propylene carbonate (PC) or other carbonate solvents. The voltage windows of MXenes can possibly be extended by introducing multiple cation species as demonstrated with model protic, lithium-containing solvents. The voltage window extension is the most pronounced for $\rm{M_2CO_2}$ early-transition-metal MXenes. Though these MXene compositions are often not considered for lithium-ion battery electrode applications due to relatively slow ion intercalation kinetics, these results suggest that some of the early-transition-metal MXenes such as $\rm{Ti_2CO_2}$ would still be excellent pseudocapacitor electrodes. The relative positions and sizes of voltage windows for the MXene compositions could assist the selection of MXenes for hybrid and asymmetric supercapacitor devices. The Zr and Hf M$_2$C MXenes may be more suitable in hybrid supercapacitor devices, while compositions with more psuedocapacitor-like responses may be applicable to asymmetric supercapacitor devices. 

We have determined that the valence of the transition metals in MXenes correlates with the energy storage. Faradaic ion adsorption induces larger ranges of oxidation states of group-VI MXenes compared to early-transition-metal compositions such as $\rm{Ti_3C_2O_2}$. We confirm that the oxidation state of the transition metals in MXene electrodes is strongly related to the pseudocapacitance, but we also show that the range of oxidation states induced in the transition metals yields larger pseudocapacitive windows as predicted by voltage-dependent cluster expansion models. The relative ranges in Bader charges serve as a computationally tractable and readily accessible descriptor for predicting MXene compositions with large pseudocapacitive voltage windows. On the basis of these trends, group-VI alloy MXenes should have large pseudocapacitive (lithium- or hydrogen-mediated) energy storage densities. Additionally, the introduction of hydrogen adsorbates to lithiated electrodes surfaces leads to an increased hybridization of the lithium \textit{s}-orbitals with the transition-metal \textit{d}-orbitals. Not only is the active lithium adsorption window extended by the introduction of hydrogen ions, but the transition-metal oxidation state is modulated more strongly compared to aprotic environments. These predictions suggest that improved pseudocapacitance can be achieved with MXene electrodes in solvents containing multiple ions that participate in Faradaic charging.

\section{Supporting Information}
Isocontours of bound charge densities and electrolyte densities near MXene electrodes, ion adsorption convex hull as a function of voltage, pseudocapacitive energy densities for other MXene compositions, Bader charge ranges for M$_3$C$_2$O$_2$ MXenes, electronic-structure changes of MXenes upon adsorption of multiple ions, cross-validation scores for cluster expansion models, predicted areal pseudocapacitances, predicted adsorption isotherms for all MXene compositions

\section{Acknowledgements}
J. G. and I. D. acknowledge primary financial support from the U.S. Department of Energy, Office of Science, Basic Energy Sciences, CPIMS Program, under Award No. DE-SC0018646.
F. V. and N. K. acknowledge financial support from Murata Manufacturing Co., Ltd. and the Center for Dielectrics and Piezoelectrics.
A portion of this work was performed by N. K. under the auspices of the DOE by Lawrence Livermore National Laboratory (LLNL) under Contract No. DE-AC52-07NA27344.
J. G. acknowledges support and training provided by the Computational Materials Education and Training (CoMET) NSF Research Training program under grant number DGE-1449785.
Calculations for this research were performed on Roar supercomputer of the Institute for Computational and Data Sciences at the Pennsylvania State University.

\clearpage
\balance

\bibliography{main}
\end{document}